\documentclass[aps,showpacs,preprint]{revtex4-1}
\usepackage{graphicx}
\usepackage{dcolumn}
\usepackage{amsmath}
\usepackage{array}
\usepackage{bm}
\usepackage{amssymb}
\usepackage{amsfonts}

\begin{document}

\title{Two- and three-body calculations within the dominantly orbital state method}

\author{Claude \surname{Semay}}
\email[E-mail: ]{claude.semay@umons.ac.be}
\author{Fabien \surname{Buisseret}}
\email[E-mail: ]{fabien.buisseret@umons.ac.be}
\altaffiliation{Haute \'Ecole Louvain en Hainaut (HELHa), Chauss\'ee de Binche 159, 7000 Mons, Belgium}
\affiliation{Service de Physique Nucl\'{e}aire et Subnucl\'{e}aire,
Universit\'{e} de Mons, UMONS Research Institute for Complex Systems,
Place du Parc 20, 7000 Mons, Belgium}

\date{\today}

\begin{abstract}
The dominantly orbital state method allows a semiclassical description of quantum systems.  At the origin, it was developed for two-body relativistic systems. Here, the method is extended to treat two-body Hamiltonians and systems with three identical particles, in $D\ge 2$ dimensions, with arbitrary kinetic energy and potential. This method is very easy to implement and can be used in a large variety of fields. Results are expected to be reliable for large values of the orbital angular momentum and small radial excitations, but information about the whole spectrum can also be obtained in some very specific cases.
\end{abstract}

\maketitle

\section{Introduction}
\label{sec:intro} 

The dominantly orbital state (DOS) method is a technique to compute approximate solutions for quantum eigenvalue problems \cite{olss97}. The first step is to compute the energy associated with a classical circular motion and then quantize the orbital angular momentum. This kind of semiclassical approximation is expected to be valid only for high values of this quantum number. Then, the radial motion is quantized around the physical circular orbit. This implies that the radial excitation must be small compared to the orbital excitation. As it will be shown, this method is very simple to implement. The only mathematical difficulty is the computation of the physical radius which is the solution of a transcendental equation. In favorable cases, the solution can be analytical. 

At the origin, the method has been developed for two-body relativistic Hamiltonians \cite{olss97} and has been used to study hadronic systems \cite{olss97,silv98,brau00}. The purpose of this work is to generalize this technique to treat two-body Hamiltonians and systems with three identical particles, in $D\ge 2$ dimensions, with arbitrary kinetic energy and potential. This is motivated by the existence of non-standard kinetic energies in some physical problems, for instance in atomic physics  with non-parabolic dispersion relation \cite{arie92}, in hadronic physics  with particle masses depending on the relative momentum \cite{szcz96}, or in quantum mechanics with a minimal length \cite{kemp95,brau99,ques10,buis10b}. Moreover, problems in $D$ dimensions can appear in various physical situations. In particular, $D=2$ systems can be used as toy models for $D=3$ systems \cite{tepe99} or are the natural framework for the physics of anyons \cite{murt91,khar05}. So, the possible domains of interest for the method are numerous. Three-body calculations are much more involved than two-body ones. The purpose of the extension of the DOS method to this kind of calculations is not to compete with numerical methods such that expansions in harmonic oscillator basis \cite{silv85} or in Gaussian states \cite{silv07}, but to provide rapidly approximate, and sometimes analytical, results.

The eigenvalues obtained by the DOS method have no definite variational character. But with very small modifications, the equations of this method can be changed into the equations of the auxiliary field (AF) method (also known as the envelope theory) \cite{silv10,silv12,sema12,hall89,hall03} which can produce upper or lower bounds on the exact solutions. The advantage of the DOS method on the AF one, is that the DOS method breaks the strong degeneracies which are typical of the AF one. These two techniques are then complementary.

This paper is organized as follows. Some elements for the physics in $D$ dimensions are briefly reminded in Sec.~\ref{sec:HD}. The DOS method for two-body systems is generalized in Sec.~\ref{sec:2B} to $D$-dimensional cases, and it is developed for three-body systems with identical particles in Sec.~\ref{sec:3B}. In both sections, examples are given to test the validity of the method. Two possible applications of the DOS method are presented in Sec.~\ref{sec:app}. Concluding remarks are given in Sec.~\ref{sec:rem}.

\section{Kinematics in $D$ dimensions}
\label{sec:HD} 

Let $\bm p$ and $\bm r$ be conjugate variables, then the operator $\bm p^2$ in $D\ge 2$ dimensions can be written as \cite{yane94} 
\begin{equation}
\label{p2D}
\bm p^2 = p_r^2 + \frac{\hat \lambda^2}{r^2} \quad \textrm{where} \quad  p_r^2=-\left( \frac{d^2}{dr^2} +  \frac{D-1}{r}\frac{d}{dr} \right),
\end{equation}
and where the orbital operator $\hat \lambda^2$ is such that \cite{aver88}
\begin{equation}
\label{la2}
\hat \lambda^2\, Y_{l,\{\mu\}}(\Omega_D) = l(l+D-2) \,Y_{l,\{\mu\}}(\Omega_D).
\end{equation}
Functions $Y_{l,\{\mu\}}(\Omega_D)$ are hyperspherical harmonics on the $D$-sphere with spherical coordinates $\Omega_D$. $l$ is the orbital quantum number and $\{\mu\}$ are the magnetic quantum numbers. In semiclassical calculations, it is usual to replace the orbital factor $l(l+a)$ by $(l+b)^2$, the two forms being equivalent for $l\gg 1$, that is to say a domain of values for which the semiclassical approximation is expected to be relevant. In this case, we have
\begin{equation}
\label{p2Dsc}
\bm p^2 = p_r^2 + \frac{\lambda^2}{r^2} \quad \textrm{with} \quad \lambda=l+\frac{D-2}{2}.
\end{equation}

\section{Two-body systems}
\label{sec:2B} 

\subsection{General formula}
\label{sec:2Bgen} 

Let us consider the two-body Hamiltonian ($\hbar=c=1$) in the center of mass frame
\begin{equation}
\label{H2B}
H=T(p) + V(r),
\end{equation}
where $\bm r$ is the relative separation between the particles and $\bm p$ the conjugate variable ($p=|\bm p|$ and $r=|\bm r|$). More complicated Hamiltonians could be considered, with spin or angular momentum dependent operators for instance, but we keep it simple to illustrate the method. 

For a circular motion ($p_r = 0$), the semiclassical ($\lambda$ is quantized) energy is given by
\begin{equation}
\label{Epr0}
E(r)=T\left( \frac{\lambda}{r} \right) + V(r).
\end{equation}
This is a crude approximation for small values of $l$, but the situation is expected to improve when $l \gg 1$. In this case, the angular momentum of the system is much larger than $\hbar$ and the dynamics is expected to be more classical. As usual, we assume that the physical orbit is determined by a minimum energy condition. This occurs for the radius $r_0$ given by
\begin{equation}
\label{r0}
\frac{\lambda}{r_0} T'\left( \frac{\lambda}{r_0} \right) = r_0\, V'(r_0).
\end{equation}
This condition has a natural explanation in terms of a centripetal force derived from a potential \cite{sema12}. It is also a semiclassical version of the generalized virial theorem \cite{luch90}, as remarked in \cite{silv12}. The corresponding energy is $E_0=E(r_0)$. 

In order to compute the radial excitation, we quantize the radial motion $\Delta r$ around a circular orbit with a fixed radius $r_0=r-\Delta r$. This motion is controlled by the Hamiltonian
\begin{equation}
\label{dH}
\Delta H = T\left( \sqrt{p_r^2 + \frac{\lambda^2}{(r_0+\Delta r)^2}} \right) + V(r_0+\Delta r) - E_0
\end{equation}
For small values of the radial quantum number $n$, we can expand $\Delta H$ in powers of $p_r^2$ and $\Delta r$. Keeping only the first non vanishing contributions, we obtain
\begin{equation}
\label{dHho}
\Delta H \approx \frac{1}{2\,\mu}p_r^2+\frac{k}{2} \Delta r^2
\end{equation}
with
\begin{equation}
\label{dHho2}
\mu = \frac{\lambda}{r_0\,T'\left( \frac{\lambda}{r_0} \right)}  \quad \textrm{and} \quad k = \frac{\lambda}{r_0^4} \left( 2\, r_0\,T'\left( \frac{\lambda}{r_0} \right) + \lambda\,T''\left( \frac{\lambda}{r_0} \right) \right) + V''(r_0).
\end{equation}
The term in $\Delta r$ is canceled by the circular condition for the orbit. Since $l\gg 1$, the radius $r$ is expected to be large. Under this condition, we can assume that $p_r^2 \approx -d^2/dr^2$, and $\Delta H$ reduces to a one-dimensional harmonic oscillator whose eigenvalues are given by $\Delta E = \sqrt{\frac{k}{\mu}}\left(n+\frac{1}{2}\right)$. Finally, the total energy, $E=E_0+\Delta E$, can be written 
\begin{eqnarray}
&&E=T\left( \frac{\lambda}{r_0} \right) + V(r_0)+\sqrt{\frac{2}{r_0^2}\,T'\left( \frac{\lambda}{r_0} \right)^2 + \frac{\lambda}{r_0^3}\,T'\left( \frac{\lambda}{r_0} \right)\,T''\left( \frac{\lambda}{r_0} \right) + \frac{r_0}{\lambda}\,T'\left( \frac{\lambda}{r_0} \right)\,V''(r_0)}\left(n+\frac{1}{2}\right) \nonumber \\
\label{E2B}
&&\textrm{with} \quad \frac{\lambda}{r_0^2} T'\left( \frac{\lambda}{r_0} \right) = V'(r_0) \quad \textrm{and} \quad \lambda=l+\frac{D-2}{2}.
\end{eqnarray}
The solutions of this system have no variational character and are expected to be reliable only when $l\gg 1$ and $n \ll l$. These constraints are strong but, for specific situations, it is possible to get results better than expected. Moreover, complementary information about the DOS solutions can be obtained using the AF method \cite{silv10,silv12,sema12,hall89,hall03}. 

The basic idea of the AF method is to replace the Hamiltonian studied $H$ by another one $\tilde H$ which is solvable, the eigenvalues of $\tilde H$ being optimized to be as close as possible to those of $H$. Practically, $\tilde H$ is a Schr\"odinger Hamiltonian with a solvable auxiliary potential $P(x)$. The systems of equations to be solved to compute the AF solutions is given by (\ref{E2B}), but without the term proportional to $(n+1/2)$, and with $\lambda$ replaced by a global quantum number $Q$. We have $Q=2n+l+D/2$ if $P(x)=x^2$ and $Q=n+l+(D-1)/2$ if $P(x)=-1/x$. The method is detailed in \cite{sema12} for $D=3$, but the extension to arbitrary dimension is trivial \cite{yane94}. It is worth noting that the AF method is not a semiclassical one, but a full quantum procedure giving approximate eigenvalues and eigenvectors. Let us define two functions $h$ and $g$ such that 
\begin{equation}
\label{hg}
T(x) = h(x^2) \quad \textrm{and}\quad V(x) = g(P(x)).
\end{equation}
If $h''(x)$ and $g''(x)$ are both concave (convex) functions, the AF eigenvalue is an upper (lower) bound of the genuine eigenvalue. If $T(p) \propto p^2$ ($V(r) \propto P(r)$), the variational character is solely ruled by the convexity of $g(x)$ ($h(x)$). For the the same computational effort, one can obtain both DOS and AF solutions. The AF solution can be a bound while the DOS solutions brings information about the breaking of the strong degeneracy implied by the global quantum number $Q$. The relevance of this formulation is tested in Sec.~\ref{sec:2Bapp} for a linear potential with an ultrarelativistic kinematics.

\subsection{Case $D=2$ and $l=0$}
\label{sec:2Bcase} 

In the particular case $D=2$ and $l=0$, $\lambda=0$ and the system~(\ref{E2B}) is ill-defined. In principle, this is not a problem since the approximation is only valid for $l \gg 1$. Nevertheless, for some systems, it is possible to obtain a good accuracy even for small values of $l$ (see following section). If the determination of $r_0$ is analytical, a generic value of $\lambda$ can be kept through the calculations. In the final formula, one must then check that the limit $\lambda\to 0$ is relevant. If only numerical calculations are possible to fix $r_0$, it is then interesting to dispose of an alternative method to treat specifically the case $\lambda=0$.

The WKB method is a powerful one to solve eigenequations without centrifugal term \cite{grif95}. With our notation, the eigenvalue $E$ can be computed by solving the following equation
\begin{equation}
\label{WKBint}
2\int_0^{r_*} T^{-1}\left( E-V(r) \right)\, dr = a\, n+b,
\end{equation}
where $a$ and $b$ are two constants. With a vanishing orbital momentum for $D=2$ ($\lambda=0$), the classical motion runs from 0 to the turning point $r_*$ given by
\begin{equation}
\label{WKBr*}
r_*= V^{-1}\left( E-T(0) \right).
\end{equation}
We can fix $a$ and $b$ in order that the WKB solution gives the exact result for the harmonic oscillator. We have then
\begin{equation}
\label{WKBE}
\int_0^{V^{-1}\left( E-T(0) \right)} T^{-1}\left( E-V(r) \right) \, dr = \pi \left( n+\frac{1}{2} \right).
\end{equation}
Equations (\ref{E2B}) and (\ref{WKBE}) allow a determination of the energy spectrum for any value of $\lambda$.

\subsection{Examples}
\label{sec:2Bapp} 

For the $D$-dimensional nonrelativistic harmonic oscillator, $T(p)=p^2/(2\, m)$ and $V(r)=k\, r^2/2$, the system~(\ref{E2B}) gives the exact result \cite{yane94}. By construction, it is also obtained by (\ref{WKBE}) for $D=2$ and $l=0$. For the $D$-dimensional Coulomb problem, $T(p)=p^2/(2\, m)$ and $V(r)=-\alpha/r$, the system~(\ref{E2B}) gives 
\begin{equation}
\label{DOScoul}
E = -\frac{m\, \alpha^2}{2\left( l+\frac{D-2}{2} \right)^2}+ \frac{m\, \alpha^2}{\left( l+\frac{D-2}{2} \right)^3}\left(n+\frac{1}{2}\right),
\end{equation}
which are the first terms of the expansion of the exact solution for $l \gg n$ \cite{yane94}. In the case $D=2$ and $l=0$, (\ref{WKBE}) gives the exact result.

Let us now look at the following semirelativistic Hamiltonian
\begin{equation}
\label{Hmes}
H=2 \sqrt{\bm p^2} + a \, r,
\end{equation}
commonly used for the study of mesons composed of light quarks \cite{fulc94}. The eigenvalues of (\ref{Hmes}) are the masses of the system. An analytical solution of the system~(\ref{E2B}) can also be obtained in this case. As it is remarked in \cite{olss97} for the case $D=3$, a more accurate approximation is obtained by computing the square energy $E^2$ but by dropping the term in $\left(n+\frac{1}{2}\right)^2$, which is coherent with the DOS approximation. The final result is then 
\begin{equation}
\label{E2Hmes}
E^2=8\, a\left( \sqrt{2}\, n+l+\frac{D-2+\sqrt{2}}{2} \right).
\end{equation}
The Hamiltonian~(\ref{Hmes}) leads to Regge trajectories: The square mass is a linear function of $l$ and $n$ with slopes independent of $D$. The ratio between the radial and orbital slopes is $\sqrt{2}$, which is in agreement with the value $\pi/2$ found in \cite{brau00b} for $D=3$ with a Bohr-Sommerfeld quantization procedure. For $D=2$ and $l=0$, (\ref{WKBE}) gives
\begin{equation}
\label{E2HmesWKB}
E^2=4\,\pi\, a\left( n+\frac{1}{2} \right),
\end{equation}
which is in quite good agreement with (\ref{E2Hmes}). The accuracy of formula~(\ref{E2Hmes}) is tested in table~\ref{tab:E2Hmes} for $D=3$, where the DOS solutions are compared with the eigenvalues computed with the high accuracy numerical Lagrange-mesh method \cite{sema01}, and with the upper bound given by the AF method \cite{silv12}
\begin{equation}
\label{E2HmesAF}
E^2_{\textrm{AF}}=8\, a\left( 2 n+l+\frac{D}{2} \right).
\end{equation}
One can see that the agreement is very good, and much better than the AF results, even for small values of $l$ and large values of $n$. With the DOS method, the oscillator degeneracy is broken and the zero point energy lowered. But, the variational character cannot be guaranteed, even if all the values are below the exact ones in this special case. It is expected that the method be more accurate for Hamiltonians close to the harmonic oscillator one, whose spectrum is exactly reproduced.

\begin{table}[htb]
\caption{Eigenmasses $E/\sqrt{a}$ of the Hamiltonian~(\ref{Hmes}) for $D=3$. For each set of quantum numbers $\{l,n\}$, the first line is an accurate value obtained by the Lagrange-mesh method \cite{sema01}, the second line is given by (\ref{E2Hmes}) and the third line by (\ref{E2HmesAF}).
\label{tab:E2Hmes}}
\begin{tabular}{rcccc}
\hline\hline
 & $n=0$ & 1 & 2 & 3 \\
\hline
$l=0$ & 3.157 & 4.709 & 5.889 & 6.871 \\
 & 3.108 & 4.579 & 5.682 & 6.603 \\
 & 3.464 & 5.292 & 6.633 & 7.746 \\
$1$ & 4.225 & 5.457 & 6.483 & 7.375 \\
 & 4.202 & 5.382 & 6.347 & 7.183 \\
 & 4.472 & 6.000 & 7.211 & 8.246 \\
$2$ & 5.079 & 6.130 & 7.047 & 7.867 \\
 & 5.065 & 6.080 & 6.949 & 7.720 \\
 & 5.292 & 6.633 & 7.746 & 8.718 \\
$3$ & 5.811 & 6.724 & 7.577 & 8.338 \\
 & 5.801 & 6.706 & 7.502 & 8.222 \\
 & 6.000 & 7.211 & 8.246 & 9.165 \\
\hline\hline
\end{tabular}\\
\end{table}

\section{Three identical particles}
\label{sec:3B} 

\subsection{General formula}
\label{sec:3Bgen} 

Let us now consider a classical Hamiltonian for three identical particles interacting via the one-body $U$ and two-body $V$ interactions
\begin{equation}
\label{H3B}
H=\sum_{i=1}^3 T(|\bm p_i|) + \sum_{i=1}^3 U\left(|\bm r_i - \bm R|\right) + \sum_{i\le j=1}^3 V\left(|\bm r_i - \bm r_j|\right),
\end{equation}
where $\sum_{i=1}^3 \bm p_i = \bm 0$ and $\bm R = \frac{1}{3}\sum_{i=1}^3 \bm r_i$ is the center of mass position. A set of internal coordinates $(\bm x=\bm x_1=\bm r_1 - \bm r_2, \bm x_2=\bm r_2 - \bm r_3, \bm x_3=\bm R)$ can be defined together with its conjugate variables $(\bm q=\bm q_1, \bm q_2,\bm q_3)$. A one-body potential such as $U$ is sometimes used to simulate confinement in hadronic systems \cite{buis10}. As mentioned in the two-body case, more complicated Hamiltonians could be considered. 

In order to apply the DOS approximation, we will impose a severe constraint on the system: the three particles move on a circular motion around the center of mass at the apex of an equilateral triangle. This constraint can be considered as the classical counterpart of the symmetrization principle which must be applied to the wavefunction (at least for bosons). Under these conditions, we have $x=|\bm x|=|\bm r_i - \bm r_j|=\sqrt{3}\,|\bm r_i - \bm R|$ and $q=|\bm q| = |\bm p_i|$, $\forall$ $i$ and $j$. The Hamiltonian~(\ref{H3B}) can then be rewritten
\begin{equation}
\label{H3Bbis}
H_{\textrm{DOS}}=3\left[ T(q) + W(x) \right] \quad \textrm{with} \quad W(x) = U\left( \frac{x}{\sqrt{3}} \right)+ V(x).
\end{equation}
In this Hamiltonian, $\bm q^2 = q_r^2+\frac{\lambda^2}{x^2}$. 

In order to interpret correctly the quantum numbers associated with the conjugate variables $\bm x$ and $\bm q$, once quantized, let us look at the total orbital angular momentum. In the center of mass frame, it is given by $\bm L = \sum_{i=1}^3 (\bm r_i - \bm R) \times \bm p_i$, and is such that 
\begin{equation}
\label{Lint}
|\bm L| = \sqrt{3}\, |\bm x \times \bm q|.
\end{equation}
If $\Lambda$ ($\lambda$) is the orbital angular momentum associated with $\bm L$ ($\bm x \times \bm q$), we have $\Lambda=\sqrt{3}\,\lambda$. Physically, $\Lambda$ must be associated with the contributions brought by the two internal variables, and we can expect that $\Lambda = \left( l_1 + \frac{D-2}{2} \right) + \left( l_2 + \frac{D-2}{2} \right)$, where $l_1$ and $l_2$ are the orbital angular momentums associated with these variables. Finally, we identify
\begin{equation}
\label{Lla}
\lambda = \frac{1}{\sqrt{3}} \left( L+D-2 \right) \quad \textrm{with} \quad L=l_1+l_2.
\end{equation}
In a semiclassical approach, the quantification of the radial motion for the three-body system is given by the integral of the quantity $3\,p_r\,\Delta r$, where $p_r$ is the radial momentum for one particle and $\Delta r$ is the variation of the radius of the circular motion. We have $3\,p_r\,\Delta r = \sqrt{3}\, q_r\,\Delta x$, where $q_r$ and $\Delta x$ are the equivalent quantities for the variables $\bm q$ and $\bm x$. If $\nu$ ($n+1/2$) is the radial quantum factor associated with $3\,p_r\,\Delta r$ ($q_r\,\Delta x$), we have $\nu=\sqrt{3}\,(n+1/2)$. Physically, $\nu$ must be associated with the contributions brought by the two internal variables, and we can expect that $\nu = \left( n_1 + \frac{1}{2} \right) + \left( n_2 + \frac{1}{2} \right)$, where $n_1$ and $n_2$ are the radial quantum numbers associated with these variables. Finally, we identify
\begin{equation}
\label{nnu}
n+\frac{1}{2} = \frac{N+1}{\sqrt{3}} \quad \textrm{with} \quad N=n_1+n_2.
\end{equation}

Using the same procedure as in Sec.~\ref{sec:2Bgen}, we can find the DOS solution for (\ref{H3Bbis}). It is written
\begin{eqnarray}
&&\frac{E}{3}=T\left( \frac{\lambda}{x_0} \right) + W(x_0)+\sqrt{\frac{2}{x_0^2}\,T'\left( \frac{\lambda}{x_0} \right)^2 + \frac{\lambda}{x_0^3}\,T'\left( \frac{\lambda}{x_0} \right)\,T''\left( \frac{\lambda}{x_0} \right) + \frac{x_0}{\lambda}\,T'\left( \frac{\lambda}{x_0} \right)\,W''(x_0)}\,\frac{N+1}{\sqrt{3}} \nonumber \\
\label{E3B}
&&\textrm{with} \quad \frac{\lambda}{x_0^2} T'\left( \frac{\lambda}{x_0} \right) = W'(x_0) \quad \textrm{and} \quad \lambda=\frac{1}{\sqrt{3}}\left( L+D-2 \right).
\end{eqnarray}
Again, this system is expected to be valid when $L\gg 1$ and $N \ll L$. The relevance of this formulation is tested in Sec.~\ref{sec:3Bapp} for two analytical solutions of (\ref{E3B}).

If the masses of the three bodies are different, the assumption $|\bm q| = |\bm p_i|$ is no longer justified. So, the present approximation scheme should not be applied. However, if a strong mass asymmetry exists between the particles, the problem can be reduced to a two- or one-body problem thanks to an adiabatic approximation. 

\subsection{Case $D=2$ and $L=0$}
\label{sec:3Bcase} 

Again, if $D=2$ and $L=0$, the system~(\ref{E3B}) is no longer well-defined. Using, the same considerations as in Sec.~\ref{sec:2Bcase}, the WKB method can be used to treat this particular case. This finally gives
\begin{equation}
\label{WKBE3}
\int_0^{W^{-1}\left( \frac{E}{3}-T(0) \right)} T^{-1}\left( \frac{E}{3}-W(x) \right) \, dx = \frac{\pi}{\sqrt{3}} (N+1).
\end{equation}
From \cite{silv10}, it is easy to determine that the exact solution for the general nonrelativistic $D$-dimensional three-body harmonic oscillator, $T(q)=q^2/(2\, m)$, $U(x)=k\, x^2$ and $V(x)=\rho\, x^2$, is written
\begin{equation}
\label{exh03}
E = \sqrt{\frac{2}{m}(k+3\,\rho)}\, ( 2\,n_1+2\,n_2+l_1+l_2 + D ).
\end{equation}
It can be checked that (\ref{WKBE3}) gives the exact result with $D=2$, $l_1=l_2=0$, and $N=n_1+n_2$. 

\subsection{Examples}
\label{sec:3Bapp} 

For the general three-body harmonic oscillator defined in the previous section, the system~(\ref{E3B}) gives
\begin{equation}
\label{DOSh03}
E = \sqrt{\frac{2}{m}(k+3\,\rho)}\, ( 2\, N + L + D ).
\end{equation}
which is the exact result with the identification $N=n_1+n_2$ and $L=l_1+l_2$. This confirms the interpretation chosen for the quantum numbers $N$ and $L$ in Sec.~\ref{sec:3Bgen}. By construction, this result is also obtained by (\ref{WKBE3}) for $D=2$ and $L=0$.

Let us now look at the following semirelativistic Hamiltonian
\begin{equation}
\label{Hbar}
H=\sum_{i=1}^3 \sqrt{\bm p_i^2} + a \sum_{i=1}^3 |\bm r_i - \bm R| + b \sum_{i\le j=1}^3 |\bm r_i - \bm r_j|,
\end{equation}
used for the study of baryons composed of light quarks \cite{buis10}. It can be seen as the three-body generalization of (\ref{Hmes}). The eigenvalues of (\ref{Hbar}) are the masses of the system. An analytical solution of the system~(\ref{E3B}) can also be obtained in this case. As for the two-body Hamiltonian, a more accurate approximation is obtained by computing the square energy $E^2$ but by dropping the term in $(N+1)^2$. The final result is then 
\begin{equation}
\label{E2Hbar}
E^2=12\, c\left( \sqrt{2}\, N+L+D-2+\sqrt{2} \right) \quad \textrm{with} \quad c=a+\sqrt{3}\, b.
\end{equation}
By comparing this equation with (\ref{E2Hmes}), one can see that the same ratio between the Regge slopes is predicted for two- and three-body systems, as expected \cite{mart86}. For $D=2$ and $L=0$, (\ref{WKBE3}) gives
\begin{equation}
\label{E2HbarWKB}
E^2=6\,\pi\, c\, ( N+1 ),
\end{equation}
which is in quite good agreement with (\ref{E2Hbar}). 

\begin{figure}[htb]
\includegraphics*[width=0.8\textwidth]{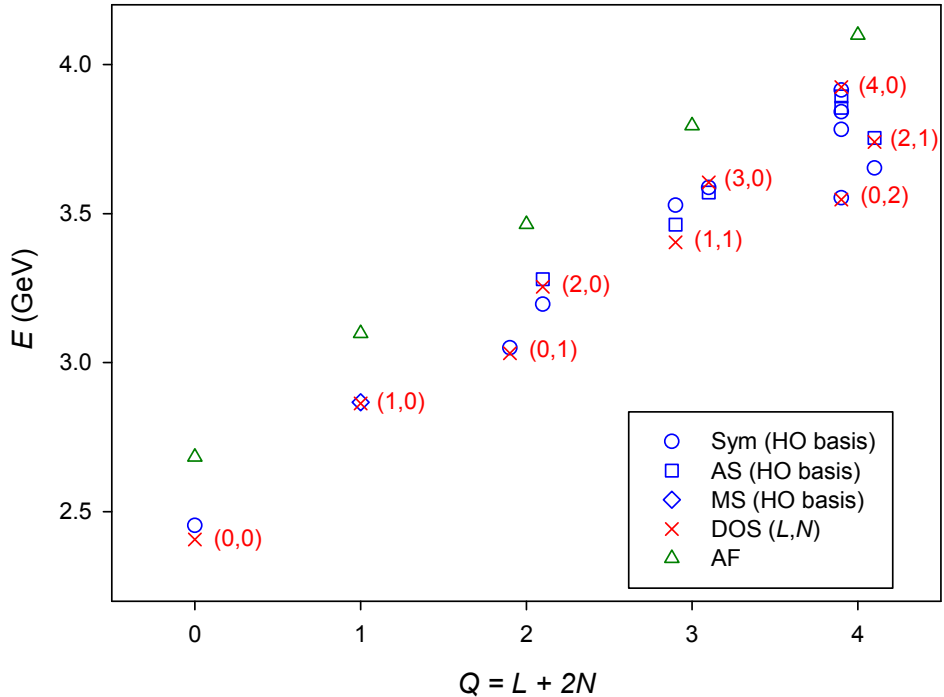}
\caption{Eigenmasses $E$ in GeV of Hamiltonian~(\ref{Hbar}) for $D=3$, and for $a=0.2$~GeV$^2$ and $b=0$, as a function of the number of quanta $Q$, obtained by an accurate computation in a harmonic oscillator basis \cite{silv85}: All values corresponding to completely symmetrical (circle) and completely antisymmetrical (square) states; Only the value for the mixed symmetry (diamond) state with $Q=1$. The corresponding approximate values (cross) given by formula~(\ref{E2Hbar}) are indicated with the quantum numbers $(L,N)$. The upper bounds of the AF method given by formula~(\ref{E2HbarAFM}) are also indicated (triangle).
\label{fig:dos3b}} 
\end{figure}

The eigenstates of the Hamiltonian~(\ref{Hbar}) have also been computed with a high accuracy method relying on the expansion of trial states on a harmonic oscillator basis \cite{silv85}. The lowest eigenstates are characterized by a large component (generally more than 90\%) spanned by one or several harmonic oscillator basis states with the same fixed value of $N=n_1+n_2$ and $L=l_1+l_2$. So they can be labeled by the number of quanta $Q=L+2\,N$, and the corresponding eigenvalues can be easily compared with the predictions of the DOS method.  All eigenvalues for the completely symmetrical and the completely antisymmetrical states are presented in Fig.~\ref{fig:dos3b}. For the clearness of the figure, only the eigenvalue corresponding to the mixed symmetry state for $Q=1$ is shown. One can see that the agreement is quite good, even if some degeneracies remain. In a pure harmonic oscillator picture, all masses associated with the same value of $Q$ should be degenerate. It can be observed in Fig.~\ref{fig:dos3b} that the DOS spectrum reproduces energy levels with any symmetry at the quantum level. Actually, $H_{\textrm{DOS}}$ and $H$ are both symmetrical for the exchange of particles, but depends on less variables. Thus, the DOS method does not have enough degrees of freedom to study quantum states with a particular symmetries and the mass degeneracies due to quantum symmetries are not all lifted. 

The AF method can gives upper bounds for the eigenvalues of Hamiltonian~(\ref{Hbar}) by means of equations very similar to (\ref{E3B}) \cite{silv10}. At the origin, this method was developed for $D=3$. But, as it relies on the exact solutions for harmonic oscillator Hamiltonians, it is not difficult to extend it to any $D$. The results is then very similar to (\ref{E2Hbar})
\begin{equation}
\label{E2HbarAFM}
E^2_{\text{AF}}=12\, c\left( 2\, N+L+D \right) =12\, c\left( Q+D \right) .
\end{equation}
As one can seen in Fig.~\ref{fig:dos3b}, it is an upper bound, but it suffers from a strong degeneracy in $Q=L+2\, N$. Formula~(\ref{E2Hbar}) does not yield bounds on eigenvalues but finally produces better results, since it has the nice feature of lifting the harmonic oscillator degeneracy. Nevertheless, the proximity between results (\ref{E2Hbar}) and (\ref{E2HbarAFM}) gives some confidence to the predictions of the DOS method. 

\section{Applications}
\label{sec:app} 

The detailed examples treated in the two previous sections are mainly relevant for the hadronic physics. We present here two different possible applications that go beyond standard Schr\"odinger equations. 

\subsection{Two-anyon systems}
\label{sec:anyon} 

In the framework of the flux tube description, a system of $N$ nonrelativistic anyons can be described by a Hamiltonian for bosons in which the following minimal prescription $\vec p_i \to \vec p_i - \vec a_i$ is achieved for each particle $i$ with \cite{khar05}
\begin{equation}
\label{ai}
\vec a_i = \alpha \sum_{j\ne i} \vec \nabla_i \theta_{ij},
\end{equation}
where $\alpha \in [0,1]$ and $\theta_{ij}$ is the relative angle between particles $i$ and $j$. $\alpha=0$ (1) corresponds to a system of $N$ bosons (fermions). This prescription leads to the presence of two-body terms ($\propto \alpha$) and three-body terms ($\propto \alpha^2$) in the Hamiltonian. 

If $N=2$, the net effect of the anyonic statistics is simply to replace angular momentum $l$ by $|l-\alpha|$ in the radial Schr\"odinger equation for the relative motion of the two particles \cite{khar05}. Approximate solutions for a two-anyon system can then be found by solving (\ref{E2B}) with $\lambda = |l-\alpha|$. The transcendental equation to be solved is then
\begin{equation}
\label{eqany}
\left| l-\alpha \right|^2 = \mu\, r_0^3\, V'(r_0),
\end{equation}
where $\mu$ is the reduced mass. Analytic solutions for $r_0$ can be found for various potentials: power law, sum of two power laws, logarithm, square root, exponential \cite{silv12}.

\subsection{Three-body systems with a minimal length}
\label{sec:minil} 

In quantum mechanics with a minimal length, the usual commutation relations between the position and momentum can be
modified from the canonical one to \cite{kemp95,brau99,ques10,buis10b}
\begin{eqnarray}\label{modcom}
[\hat x^a, \hat p^b]&=& i(\delta^{ab}(1+\beta\,\hat p^2)+2\beta\,\hat p^a \hat p^b), \\ \nonumber
 [\hat x^a, \hat x^b ]&=&[\hat p^a, \hat p^b]=0,
\end{eqnarray} 
provided that the deformation parameter $\beta$ is small enough to work at first order in $\beta$. Note that $a,b=1,\dots,D$. When $D>1$, a representation of the above algebra is $\hat x^a=x^a$ and $\hat p^a=(1+\beta p^2) p^a$, where $x^a$ and $p^a$ satisfy the undeformed Heisenberg algebra \cite{brau99}. So for nonrelativistic problems, one can use the following kinetic term for a particle with a mass $m$,
\begin{equation}
\label{minx}
T(p)=\frac{p\,^2}{2 m} + \frac{\beta}{m}  p\,^4.
\end{equation}

One readily sees that the DOS approximation can then be used to compute energy spectra of two- and three-body Hamiltonians with a minimal length. Let us give here the DOS result for a three-particle system bound by two-body harmonic potentials $V(x)=k\, x^2$. At the first order in $\beta$, Eq.~(\ref{E3B}) tells us that 
\begin{equation}
x_0=\lambda^{1/2}(2km)^{-1/4}+\lambda^{3/2}(2km)^{1/4}\beta ,
\end{equation}
and that 
\begin{equation}\label{HOmin}
E =\sqrt{\frac{6k}{m}}(2N+L+D)+2k (L+D-2)(6N+L+D+4)\beta.
\end{equation}
To our knowledge, it is the first time that this spectrum is computed. Hence the DOS method is general enough to be applied to a wide range of quantum mechanical problems. Note that formula (\ref{HOmin}) is exact when $\beta=0$, while quadratic corrections in the quantum numbers have already been found at order $\beta$ in the two-body case for $D=3$ \cite{brau99}.

\section{Concluding remarks}
\label{sec:rem} 

The dominantly orbital state (DOS) method is a semiclassical technique to compute approximate solutions for quantum eigenvalue problems. Developed at the origin for two-body relativistic systems \cite{olss97}, it is extended here to treat two-body Hamiltonians and systems with three identical particles, in $D\ge 2$ dimensions, with arbitrary kinetic energy and potential. Given the nature of the DOS method, it seems not possible to apply to other many-body systems. It is expected to be valid only for high values of the orbital angular momentum and small radial excitations, but it can give good results for the whole spectra in some particular cases. This method is very simple to implement and needs only the solution of an algebraic system of equations very similar to the one for the auxiliary field (AF) method \cite{silv10,silv12,sema12,hall89,hall03}. In favorable cases, analytical formulae can be obtained. The two methods need the same computational effort and are complementary: The AF one can produce upper or lower bounds, depending on the Hamiltonian, while the DOS one breaks the degeneracy inherent to the AF technique. 

A lot of accurate techniques exist to solve numerically two-body problems. So, both DOS and AF methods are specially interesting for cases in which they can produce an analytical solution. They can be added to the arsenal of other methods able to produce analytical results: The expansion in a small number of basis states; the WKB method, specially for vanishing angular momentum and high radial excitations; the perturbation theory in very specific cases. The peculiarity of the DOS method is that the eigenvalues obtained are expected to be reliable at least for high angular momentum and small radial excitations. 

A three-body problem is always difficult to solve, and techniques to implement it can be very heavy \cite{silv85,silv07}. In this case, the purpose of the DOS method is not to compete with these very accurate techniques, but instead to provide rapidly approximate, and sometimes analytical, results. These results are expected to be valid at least in the high angular momentum and small radial excitations regimes. Despite the strong constraint imposed on the motion of the particles, the results obtained seem reliable: We reproduce the harmonic spectrum exactly and the Regge trajectories of relativistic systems with a linear confinement. This property can be particularly interesting when applied to hadronic physics, for example \cite{buis13}. 

\section*{Acknowledgments}

C.S. would thank Gwendolyn Lacroix for a useful suggestion.

\end{document}